\documentclass[twocolumn,showpacs,preprintnumbers,amsmath,amssymb,nofootinbib]{revtex4}

\usepackage{graphicx,epsf, epsfig, psfig, amssymb}
\usepackage{bm} 

\def\be{\begin{equation}}
\def\ee{\end{equation}}
\def\beq{\begin{eqnarray}}
\def\eeq{\end{eqnarray}}

\def\ii{{\rm i}}

\def\IL{\relax{\rm I\kern-.18em L}}

\def\f{\frac}

\begin{document}

\title{Acoustic black holes
}

\author{Vitor Cardoso}
\email{vcardoso@wugrav.wustl.edu} \affiliation{McDonnell Center for
the Space Sciences,
Department of Physics, Washington University, St. Louis, Missouri 63130, USA\\
                and \\
Centro de F\'{\i}sica Computacional, Universidade de Coimbra,
P-3004-516 Coimbra, Portugal}

\date{\today}

\begin{abstract}
We discuss some general aspects of acoustic black holes. We begin by
describing the associated formalism with which acoustic black holes
are established, then we show how to model arbitrary geometries by
using a de Laval nozzle. It is argued that even though the Hawking
temperature of these black holes is too low to be detected, acoustic
black holes have interesting classical properties, some of which are
outlined here, that should be explored.
\end{abstract}


\maketitle

\section{Introduction}
Black holes are among the most fascinating objects in physics. The
fact that they are pure objects, in the sense that they are made
from spacetime itself, explains why they have taken such a special
place in general relativity. There is a powerful and elegant
mathematical machinery to describe them \cite{chandra,ruffini}, and
their classical and quantum properties are well understood within
the general relativity framework. In this setting, the properties of
isolated black holes have been thoroughly investigated. The much
more complex processes that take part in the surroundings of
astrophysical black holes, the interaction of black holes with
matter (accretion disks, magnetic fields, etc) \cite{shapiro} or
even with other black holes (for example, the problem of the head-on
collision of two black holes is now solved \cite{gleiser}) are more
or less well understood as well. On the semi-classical side, Hawking
\cite{hawking} showed that when quantum effects are taken into
account, black holes are not really black: they slowly evaporate by
emitting an almost thermal radiation. Hawking's prediction has been
theoretically confirmed time and again in very different ways.  The
discovery of Hawking radiation uncovered a number of fundamental
questions: among them the information puzzle, the issue of the black
hole final state, and so on. Some of these issues can be tackled
only in a more fundamental theory, because classical general
relativity is not the ultimate theory of gravity since it does not
embody the principles of quantum mechanics. A consistent theory of
quantum gravity requires a modification of classical general
relativity, and in the two alternative theories more in fashion
nowadays, string theory and loop quantum gravity, black holes still
occupy a special position. String theory's charm, for instance,
derives in part from a couple of remarkable breakthroughs in
connection with black hole physics, namely the entropy calculation
by a counting of micro states, and the computation of greybody
factors \cite{vafa,zwie,oz,malda}.

The progress in understanding black holes has been immense, over
these last forty years since their concept was born, and they now
play a central role in modern physics. Despite this, the lack of
experimental tests has always been a drawback, for general
relativists, and for people studying black holes in particular. An
important step to make black holes more accessible (from an
experimental point of view) was given in 1981 by Unruh \cite{unruh},
who came up with the notion of analogue black holes. While not
carrying information about Einstein's equations, the analogue black
holes devised by Unruh do have a very important feature that defines
black holes: the existence of an event horizon. The basic idea
behind these analogue acoustic black holes is very simple: consider
a fluid moving with a space-dependent velocity, for example water
flowing throw a variable-section tube. Suppose the water flows in
the direction where the tube gets narrower. Then the fluid velocity
increases downstream, and there will be a point where the fluid
velocity exceeds the local sound velocity, in a certain frame. At
this point, in that frame, we get the equivalent of an apparent
horizon for sound waves. In fact, no (sonic) information generated
downstream of this point can ever reach upstream (for the velocity
of any perturbation is always directed downstream, as a simple
velocity addition shows). This is the acoustic analogue of a black
hole, or a {\it dumb hole}. These objects share more properties with
true, gravitational black holes, besides the existence of horizons:
they display geodesics, wave effects in their vicinity and, as we
shall see they also emit Hawking radiation. Nevertheless they are
not true black holes, because the acoustic metric satisfies the
equations of fluid dynamics and not Einstein's equations. One
usually expresses this by saying that they are analogs {\it of}
general relativity, because they provide an effective metric and so
generate the basic kinematical background in which general
relativity resides. They are not models {\it for} general
relativity, because the metric is not dynamically dependent on
something like Einstein's equations \cite{visser,novello}. Following
on Unruh's dumb hole proposal many different kinds of analogue black
holes have been devised, based on condensed matter physics, slow
light etcetera \cite{visser,novello,tapas}. Analogue black holes
have been the subject of intense study because of the Hawking
radiation they emit. In fact, it is now clear that the appearance of
Hawking radiation does not depend on the dynamics of the Einstein
equations, but only on their kinematical structure, and more
specifically on the existence of an apparent horizon
\cite{novello,visserhawking}. The experimental verification of the
Hawking effect is not easy though. Unfortunately astrophysical black
holes, having a Hawking temperature much smaller than the
temperature of the cosmic microwave background, accrete matter more
efficiently than they evaporate. However, since Hawking radiation
crucially depends on the existence of an apparent horizon, the
analogues just described do emit Hawking radiation, and this was and
still is the primary reason to study them. At present the Hawking
temperatures associated to these analogues are too low to be
detectable, but the situation is likely to change in the near future
(see for instance \cite{unruh2}).

The importance of classical properties of analogue black holes have
been somewhat underestimated. First, even though building (Hawking)
very hot analogue black holes may be extremely difficult, building
them {\it any} acoustic black hole is not. Thus we can easily have
an acoustic black hole in almost any lab. What good are these black
holes for?

(i) They have an horizon. Hawking radiation is not the only
interesting thing going on when an event horizon shows up! In
particular, I would say that the ``only in-going waves at the
horizon'' boundary condition would be interesting to see
experimentally, with all its associated phenomena, some of which are
mentioned below.

(ii) Geodesics. This is a particularly interesting application. As
will be shown one can easily mimic several geometries simply by
varying the cross section of a de Laval nozzle. Thus we can observe
the geodesics in different spacetimes easily.

(iii) Measuring absorption cross-sections. This would also be an
interesting application of analogue black holes, to measure
absorption cross-sections, glory effects, etc, and compare them with
theoretical predictions \cite{matzner}.

(iv) Superradiance. This phenomenon, involving rotating black holes
\cite{zel}, was in the basis for the discovery of Hawking radiation
\cite{israel}. To our knowledge this effect was never experimentally
verified (not including Cherenkov radiation in this category
\cite{bekenstein}), but it should not be very hard to reproduce in
the lab using acoustic black holes
\cite{waveanalog1,waveanalog2,savage}.

(v) Quasinormal modes. The resonance modes of black holes, called
quasinormal modes (QNMs) are a very important concept in any
discussion involving gravitational radiation by black holes, the
approach to equilibrium and black hole detection
\cite{kokkotas,vitorthesis2004}. The QNMs of black holes use the
in-going waves at the horizon boundary condition, and they usually
have quite an interesting spectra of frequencies
\cite{cardososhijunlemos}. The QNMs of some analogue black holes
have already been computed
\cite{waveanalog1,waveanalog2,QNManalogue}.

(vi) Late-time tails. Black holes have no hair, and it is lost at
late times as a power-law falloff \cite{tails}. Late-time tails can
also be studied using black hole analogues \cite{waveanalog1}.

(vii) Analog black branes. It should be rather easy to implement
other analogue black objects, such as black branes and strings,
which could deepen our understanding about these geometries.

(viii) Interaction of black holes with electric and magnetic fields.
On a more speculative vein, it is even possible in principle to
simulate in the lab the interaction of astrophysical black holes
with matter and with electromagnetic fields. It is even possible
that one might be able to study effects such as the Blandford-Zjanek
process \cite{shapiro,blandford}. This would be a tremendous
motivation to use and explore  analogue black holes. Some steps
along this direction, although not directly connected to analogue
black holes, were given in \cite{sisan}.

These are just classical aspects of black holes, but even these {\it
must} be mastered before embarking on experimental Hawking radiation
detection. Not only must one control what happens in the
experimental situation, but the understanding of classical phenomena
may bring clues on how to favor the probabilities to detect Hawking
radiation. It is also worth stressing that some purely classical
phenomena shed light on quantum aspects of (analogue and
general-relativistic) black hole physics. For example, positive and
negative norm mixing at the horizon leads to non-trivial Bogoliubov
coefficients in the calculations of Hawking radiation
\cite{corleyjacobson}; superradiant instabilities of the Kerr metric
are related to the quantum process of Schwinger pair production
\cite{schwinger,detweiler,zouros,furu}; and more speculatively
(classical) highly damped black hole oscillations could be related
to area quantization \cite{hod} (this possibility was discarded in
\cite{waveanalog1} because of the failure to satisfy the laws of
black hole mechanics \cite{visserlaws}, but the situation may change
\cite{cadoni}).

Most of what has been said applies equally well to other types of
analogue black holes (see for instance \cite{novello,schutzhold}),
but for simplicity we shall here deal only with acoustic black
holes. Some aspects of acoustic black holes will be explored: we'll
explain how to generate a large class of acoustic black holes by
using a de Laval nozzle with a variable cross section profile. We
will see how to make a simple black brane and study its stability
properties. We will then explain why in certain situation the
analogue branes are unstable \cite{braneinstability}. We will make a
small review of what has been done so far concerning classical
aspects of wave propagation in acoustic black holes, taking the
($2+1$)-dimensional rotating black hole as a case study.

\section{Effective acoustic geometry}
This section will be as self contained as possible, because we want
to make explicit the assumptions that go with the usual derivation
of the acoustic metric. However, this derivation can be found in the
monograph by Matt Visser \cite{visser}. Let us start with the
equations of fluid dynamics, and try to arrange them in such a way
that an effective metric stands out naturally. The fundamental
equations of fluid dynamics \cite{booklandau,bookcomp,chandrabookf}
are the equation of continuity \be
\partial _t + \nabla .( \rho {\bm v})=0\,, \label{continuity}\ee and
Euler's equation \be \rho \frac{{\bm v}}{dt}\equiv \rho [\partial _t
{\bm v}+({\bm v}.\nabla){\bm v}]=-\nabla p+{\bm F}\,,\label{euler}
\ee where ${\bm F}$ are for the moment all the external forces
acting on the fluid. Hereafter we shall make the following
assumptions: (i) the external forces are all gradient-derived, or
${\bm F}=-\rho\nabla \Phi$. Thus we are neglecting viscosity terms
in Navier-Stokes equation. (ii) the fluid is locally irrotational,
and introduce the velocity potential $\psi$, ${\bm v}=-\nabla \psi$;
and (iii) the fluid is barotropic, i.e., the density $\rho$ is a
function of pressure $p$ only. In this case, we can define \be
h(p)=\int_0^p \frac{dp'}{\rho(p')}\,,\ee or \be \nabla
h=\frac{\nabla p}{\rho}\,.\ee Euler's equation can be written as \be
\partial _t \psi +h +\frac{1}{2}(\nabla \psi)^2+\Phi=0\,.\ee

To study sound waves, we follow the usual procedure and linearize
the continuity and Euler's equations around some background flow, by
setting $\rho= \rho _0+\epsilon \rho _1\,,\,\,p=p_0+\epsilon
p_1\,,\,\,\psi=\psi_0+\epsilon \psi _1$, and discarding all terms of
order $\epsilon ^2$ or higher. The external potential $\Phi$ is
taken as constant. The continuity equation yields \beq \partial _t
\rho _0 +\nabla .(\rho _0 {\bm v_0})=0 \,, \\ \partial _t \rho _1
+\nabla .(\rho _1 {\bm v_0}+\rho _0{\bm v_1})=0
\,.\label{contlinea}\eeq Linearizing the enthalpy we get
$h(p_0+\epsilon p_1)\sim h(p_0)+\epsilon
\frac{dh}{dp}_{p=p_0}=h_0+\epsilon \frac{p_1}{\rho _0}$. Inserting
this in Euler's equation one gets \beq -\partial _t \psi
_0+h_0+\frac{1}{2}(\bigtriangledown \psi _0)^2+\Phi=0\,,\\
p_1=\rho _0(\partial _t \psi _1+{\bf v_0}.\nabla \psi _1)=\,.
\label{p1} \eeq On the other hand, since the fluid is barotropic we
have \be\rho _1=\frac{\partial \rho}{\partial p}p_1\,,\ee and using
(\ref{p1}) this is the same as \be \rho _1 =\frac{\partial
\rho}{\partial p}\rho _0(\partial _t \psi _1+{\bf v_0}.\nabla \psi
_1)\,.\ee Finally, substituting this into (\ref{contlinea}) we get
\beq 0=-\partial _t \left ( \frac{\partial \rho}{\partial p}\rho
_0(\partial _t \psi _1+{\bf v_0}.\nabla \psi _1)\right )+ \\
\nabla. \left(\rho _0 \nabla \psi _1-\frac{\partial \rho}{\partial
p}\rho_0 {\bm v_0}(\partial _t \psi _1+{\bf v_0}.\nabla \psi _1)
\right )\,.\eeq It can now easily be shown \cite{visser} that this
equation can also be obtained from the usual curved space
Klein-Gordon equation \be
\partial _{\mu} (\sqrt{-g}g^{\mu \nu}\partial _{\nu}\psi)=0\,,\ee
with the effective metric $g^{\mu \nu}$ given by
\begin{eqnarray}
g^{u \nu}\equiv \frac{1}{\rho _0 c} \left[
\begin{array}{ccc}
-1 &\vdots&-v_0^{j}\\
\hdots \hdots &.&\hdots \hdots \hdots\\
-v_0^{i}&\vdots& (c^2 \delta _{ij}-v_0^iv_0^j)
\end{array}
\right]\,. \label{metricvisserinv}
\end{eqnarray}
Neither of the background quantities is assumed constant through the
flow, and so they are in general dependent on the coordinates along
the flow. Here we have used the definition of the local sound speed
$c^{-2}=\frac{\partial \rho}{\partial p}$. We can see that the
propagation of sound waves in a fluid is equivalent to the
propagation of a scalar field in a generic curved spacetime
described by (\ref{metricvisserinv}), or in covariant form by
\begin{eqnarray}
g_{u \nu}\equiv \frac{\rho _0}{c} \left[
\begin{array}{ccc}
-(c^2-v_0^2) &\vdots&-v_0^{j}\\
\hdots \hdots\hdots &.&\hdots \hdots\\
-v_0^{i}&\vdots&\delta _{ij}
\end{array}
\right] \label{metricvisser}
\end{eqnarray}
This means that {\it all} properties of wave propagation in curved
space hold also for the propagation of sound waves. The power of
this effective geometry should be clear: first, by changing the
background flow, we change the effective acoustic metric. Second,
since this geometry clearly has an apparent horizon at the point
where $v_0=c$, and since the existence of an apparent horizon
implies Hawking radiation, then there should be Hawking radiation in
this geometry, which takes the form of phonons \cite{unruh}. The
Hawking temperature can be computed to yield \cite{visser}: \be
kT_H=\frac{\hbar }{2\pi}\frac{\partial(c-v_\bot)}{\partial n}
\label{hawkinhtemp}\,,\ee where $v_\bot$ is the component of the
fluid velocity normal to the horizon, and ${\bm n}$ is the unit
vector normal to the horizon. This can also be written as \be
T_H=1.2 \times 10^{-9}{\rm K}\,{\rm m}\left[ \frac{c}{1000\,{\rm
ms^{-1}}}\right]\left[ \frac{1}{c}\frac{\partial(c-v_\bot)}{\partial
n}\right ] \label{temp2}\ee

This is, for all practical purposes a number too low to be detected,
and it is even more so if one notices that one has to deal with the
ambient noise. Despite the fact that one cannot observe Hawking
radiation, one can still measure classical aspects of black holes.
So we now turn to this, but first we explain how we can mimic
several geometries by using a de Laval nozzle.
\section{Shaping the nozzle}

\subsection{The de Laval nozzle}
\begin{figure}
\centerline{\includegraphics[width=9 cm,height=3 cm] {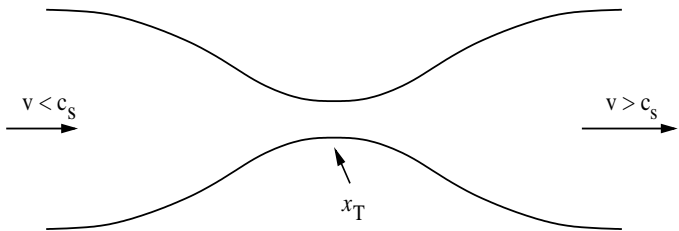}}
\caption{A sketch of a de Laval nozzle, used to make a smooth
transition from subsonic to supersonic flow. The velocity of the
fluid $v(x)$ equals the local velocity of sound $c(x)$ at the throat
of the nozzle, $x=x_{T}$. The cross-section at this point is denoted
by $A_T$. } \label{fig:nozzle}
\end{figure}

A de Laval nozzle is a device which can be used to accelerate a
fluid up to supersonic velocities. They were first used in steam
turbines, but they find many applications in rocket engines, nozzles
in supersonic wind tunnels, etc. It consists of a converging pipe,
where the fluid is accelerated, followed by a throat which is the
narrowest part of the tube and where the flow undergoes a sonic
transition, and finally a diverging pipe where the fluid continues
to accelerate. It is sketched in Fig. \ref{fig:nozzle}.
\vskip 1mm Consider now a steady, isentropic flow through the
nozzle, which has a varying cross-section $A(x)$, where $x$ is the
arc length along a streamline. Logarithmic differentiation of the
continuity equation \be \rho v A=\frac{dm}{dt}={\rm const}\,,
\label{continuityeq} \ee yields \be
\frac{1}{v}\frac{dv}{dx}+\frac{1}{A}\frac{dA}{dx}+
\frac{1}{\rho}\frac{d \rho}{dx} =0 \,. \label{continuitylogdif} \ee
For isentropic flow $p=p(\rho)$ and the definition for the speed of
sound $\left [c^2=\frac{\partial p}{\partial \rho}\right ]_{\rm
const entropy}$ immediately gives  $c^2=\frac{d p}{d \rho}$. Using
this in (\ref{continuitylogdif}) we obtain \be
\frac{1}{v}\frac{dv}{dx}+\frac{1}{A}\frac{dA}{dx}+ \frac{1}{c^2
\rho}\frac{d p}{dx} =0 \,. \label{continuitylogdif2} \ee Using the
component of Euler's equation along the streamline \be \rho v
\frac{d v}{d x}=-\frac{dp}{dx}\,, \label{Euler} \ee and combining it
with equation (\ref{continuitylogdif2}) we have finally \be
\frac{1}{v}(1-\frac{v^2}{c^2}) \frac{dv}{dx}=
-\frac{1}{A}\frac{dA}{dx}  \,. \label{Nozzle} \ee According to this
differential equation, when the flow is subsonic ($v<c$), $dv/dx$
and $dA/dx$ have opposite signs. So narrowing the pipe will make the
gas flow faster, which is what we expect from common experience. In
fact, for very small $v$, equation (\ref{Nozzle}) can be written as
$dv/v=-dA/A$ and thus $vA$ is a constant, a well known result for
incompressible fluids. The situation is opposite for $v>c$, when
$dv/dx$ and $dA/dx$ have the same sign. This means that a region of
increasing cross- section will accelerate the flow. The nozzle
equation (\ref{Nozzle}) shows that the transonic flow through the
nozzle must reach $v=c$ at the throat where $dA/dx=0$. This is a
necessary condition, but not a sufficient one. Whether the actual
flow will be transonic depends of course on the lower boundary
condition, i.e., the velocity $v_0$ at the entrance of the nozzle.
For a given profile $A(x)$ of the nozzle $v_0$ has to have exactly
the right value: if $v_0$ is too small the flow will remain subsonic
everywhere. If $v_0$ is too large the velocity will reach $v=c$
upstream from the throat, at $x<x_T$, where $dA/dx<0$ and so $dv/dx
\rightarrow -\infty$ at that point. The flow will stagnate which
results in a shock between the flow and the low velocity region. So
in this case there is no smooth transonic flow.

I shall from now on assume that the boundary conditions are such
that there is a smooth transition from sub to supersonic flow, at
the throat located at $x=x_T$. Given a nozzle profile $A(x)$ and an
equation of state, then Eq. (\ref {Nozzle}) together with
(\ref{metricvisser}) describes completely, for $x<x_T$, an acoustic
black hole (I note that the equation of state $p=p(\rho)$ will
allow, from Eq. (\ref{continuitylogdif}), to have $c$ as a function
of $v$). Conversely, given an equation of state, a whole family of
acoustic black holes can be obtained, by simply varying $A(x)$. For
example, for a perfect gas, we have \cite{booklandau}
$c^2=c_0^2-(\gamma- 1)/2 v^2$, where $c_0$ is the speed of sound at
the location corresponding to $v=0$, and $\gamma$ is the ration of
specific heats. Now that we have $c$ as  function of $v$, Eq.
(\ref{Nozzle}) allows one to specify the velocity profile and
therefore the full metric (\ref{metricvisser}) as a function of
$A(x)$.

\subsection{Acoustic black holes by different nozzle configurations}
Let us now suppose that the dependence of $c$ on $x$ is small, i.e.,
let us assume that $\partial c/\partial x $ is not very large (as
happens for example for perfect gases), and therefore that $c \sim
{\rm const}$. With the assumption of constant sound velocity,
equation (\ref{Nozzle}) can be solved yielding \be
\frac{1}{v}e^{v^2/(2c^2)}=A(x)C\,, \label{sol} \ee where the
constant $C$ is found by applying the condition at the throat $v
(x_T)=c$. This gives \be
\frac{1}{v}e^{v^2/(2c^2)-1/2}=A(x)/A(x_T)\,. \label{sol2} \ee Let us
choose the following generic form for $A$: \be
A(x)=\frac{1}{f(x)}e^{\left[A(x_T)f(x)\right]^2/2-1/2}\,.
\label{sol3} \ee For this to be a consistent solution we must have
$dA/dx=0$ at $x=x_T$, which results in the constraint \be
f'(x_T)e^{\left[A(x_T)f(x_T)\right]^2/2-1/2}\left[A(x_T)^2-\frac{1}{f(x_T)^2}
\right]=0\,. \label{Aconstr} \ee This constraint, together with the
definition (\ref{sol3}) can be satisfied if one chooses \be
f(x_T)=\frac{1}{A_T}\,. \label{Aconstr2} \ee Equation (\ref{sol2})
is then trivially solved by \be
\frac{v}{c}=f(x)A(x_T)\label{mimicg}\ee

So we conclude that in order to mimic some metric, we have to be
able to simulate the correct background flow, as indicated by Eq.
(\ref{metricvisser}). Now, to mimic the background velocity, one
only has to build a de Laval nozzle according to (\ref{mimicg}), and
we have our problem solved. One can also look at the Hawking
radiation of acoustic black holes in nozzles. This was done recently
by Barcelo, Liberati and Visser \cite{barcelo2}.


\section{Classical wave phenomena near acoustic black holes}
Some classical aspects of wave propagation in acoustic black holes
have been explored in \cite{waveanalog1,waveanalog2}. Here I will
summarize some of their results and also comment on absorption
cross-sections, focusing always on the ($2+1$)-dimensional rotating
acoustic black hole \cite{visser}, which I now describe.

A simple rotating acoustic black hole geometry was presented in
\cite{visser} modeling a ``draining bathtub'', idealized as a
($2+1$)-dimensional flow with a sink at the origin. I will show that
a simple generalization of this geometry can mimic a black brane,
and I will show why, despite recent claims \cite{braneinstability},
it is not unstable.

Consider a fluid having (background) density $\rho$.  Assume the
fluid to be locally irrotational (vorticity free), barotropic and
inviscid. From the equation of continuity, the radial component of
the fluid velocity satisfies $\rho v^r\sim 1/r$. Irrotationality
implies that the tangential component of the velocity satisfies
$v^\theta\sim 1/r$. By conservation of angular momentum we have
$\rho v^\theta\sim 1/r$, so that the background density of the fluid
$\rho$ is constant. In turn, this means that the background pressure
$p$ and the speed of sound $c$ are constants. The acoustic metric
describing the propagation of sound waves in this ``draining
bathtub'' fluid flow is \cite{visser}:
\beq ds^2&=& -\left (c^2-\frac{A^2+B^2}{r^2} \right
)dt^2+\frac{2A}{r}drdt-\nonumber \\
& & 2Bd\phi dt+dr^2+ r^2d\phi^2\,. \label{metric1} \eeq Here $A$ and
$B$ are arbitrary real positive constants related to the radial and
angular components of the background fluid velocity:
\be {\vec v}=\frac{-A {\hat r}+B {\hat \theta}}{r}\,. \ee
In the non-rotating limit $B=0$ the metric (\ref{metric1}) reduces
to a standard Painlev\'e-Gullstrand-Lema\^itre type metric
\cite{PGL}. The acoustic event horizon is located at $r_H=A/c$, and
the ergosphere forms at $r_{ES}=(A^2+B^2)^{1/2}/c$.

Some physical properties of our ``draining bathtub'' metric are more
apparent if we cast the metric in a Kerr-like form performing the
following coordinate transformation (where again we correct some
typos in \cite{basak2}): \be dt\rightarrow
d\tilde{t}+\frac{Ar}{r^2c^2-A^2}dr\,,\qquad d\phi\rightarrow
d\tilde{\phi}+\frac{BA}{r(r^2c^2-A^2)}dr\,. \label{coordtransf} \ee
Then the effective metric takes the form \beq ds^2&=&
-\left(1-\frac{A^2+B^2}{c^2r^2} \right)c^2 d\tilde{t}^2+\nonumber \\
& & \left(1-\frac{A^2}{c^2r^2} \right )^{-1}dr^2 -2B
d\tilde{\phi}d\tilde{t}+r^2d\tilde{\phi}^2\,. \label{metric2} \eeq
Notice an important difference between this acoustic metric and the
Kerr metric: in the ($t,t$) component of the metric (\ref{metric2})
the parameters $A$ and $B$ appear as a {\it sum} of squares. This
means that, {\it at least in principle}, there is no upper bound for
the rotational parameter $B$ in the acoustic black hole metric,
contrary to what happens in the Kerr geometry.

\subsection{QNMs}
\begin{figure}
\centerline{\includegraphics[width=6 cm,height=6 cm] {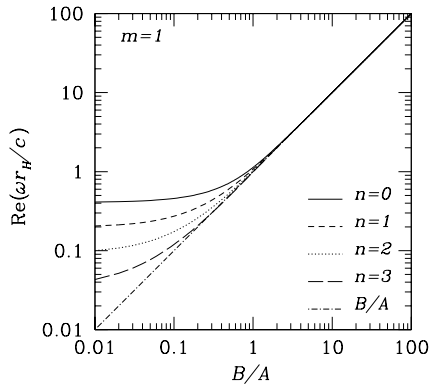}}
\caption{The real part of the QN frequency as a function of the
rotation parameter $B/A$, for several overtones of a $m=1$ mode.
Here, $r_H=A/c$ is the horizon radius. Note how all the several
lowest overtones ``coalesce'' in the high rotation regime, all
growing linearly with $B/A$.} \label{fig:f1}
\end{figure}

\begin{figure}
\centerline{\includegraphics[width=6 cm,height=6 cm] {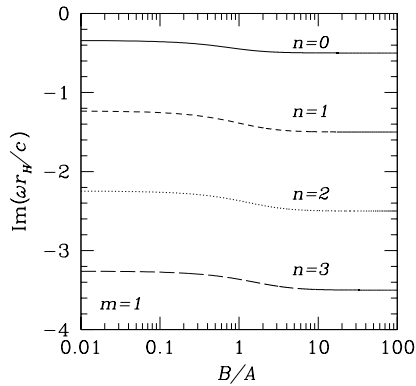}}
\caption{The imaginary part of the QN frequency as a function of the
rotation parameter $B/A$, for several overtones of a $m=1$ mode. It
is clear from this plot that the imaginary part of the QN
frequencies of $m>0$ modes is very insensitive to the rotation of
the black hole. } \label{fig:f2}
\end{figure}

\begin{figure}
\centerline{\includegraphics[width=6 cm,height=6 cm] {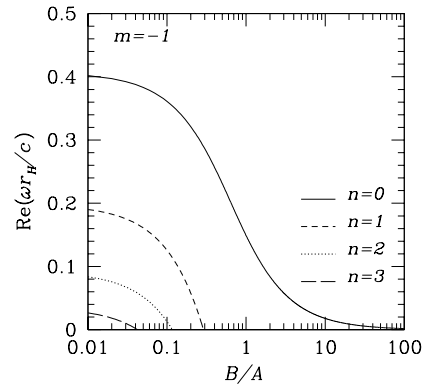}}
\caption{The real part of the QN frequency as a function of the
rotation parameter $B/A$, for several overtones of a $m=-1$ mode.
Notice that for each overtone number $n$ there is a critical
rotation at which the mode crosses the axis, i.e., there is a
critical rotation $B/A$ at which the real part of the QN frequency
is zero. Higher overtones cross the axis at a slower rotation. We
have not been able to follow the mode beyond this point.}
\label{fig:f3}
\end{figure}

\begin{figure}
\centerline{\includegraphics[width=6 cm,height=6 cm] {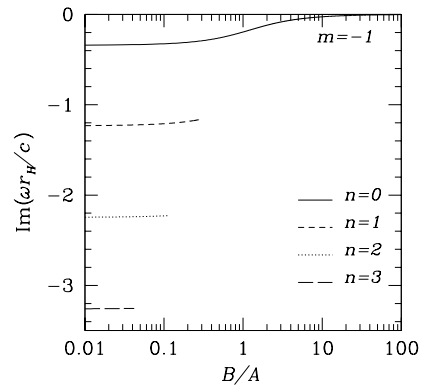}}
\caption{The imaginary part of the QN frequency as a function of the
rotation parameter $B/A$, for several overtones of a $m=-1$ mode. We
have not been able to follow the modes beyond a certain critical
point (defined as the rotation $B/A$ for which the real part of the
QN frequency is zero). Nevertheless, an judging by the modes we did
manage to follow, namely the fundamental mode, it seems that ${\rm
Im}[\omega_{QN}]$ never crosses the axis, i.e., it is always
negative, and therefore the mode is stable. } \label{fig:f4}
\end{figure}

Black holes, like so many other objects, have characteristic
oscillation or ringing modes, which are called quasinormal modes
(QNMs) \cite{kokkotas,vitorthesis2004,cardososhijunlemos}, the
associated frequencies being termed QN frequencies, or
$\omega_{QN}$. The QN frequencies of the ($2+1$)-rotating acoustic
black hole, described by the metric (\ref{metric2}) were recently
computed in \cite{waveanalog1,waveanalog1}, and so were the QNMs of
the canonical acoustic black hole. The numerical results, consistent
with a WKB analysis, are shown in Figs. \ref{fig:f1}-\ref{fig:f4}.

\vskip 1mm

\vskip 1mm

\vskip 1mm

\vskip 1mm

\noindent {\bf (i) $m>0$:} In Fig. \ref{fig:f1}-\ref{fig:f2} we show
results pertaining to perturbations having positive $m$, i.e.,
co-rotating waves. In Fig. \ref{fig:f1} we show the real part of the
QN frequencies for $m=1$ modes as a function of the black hole
rotation. Higher $m$ modes follow a similar pattern. One can see
from this plot that for low black hole rotation parameter $B$ the
different overtones are clearly distinguished, but that as the
rotation increases they tend to cluster and behave very similarly.
For very large rotation $B$, all the overtones behave in the same
manner, and in this high rotation regime the real part of the QN
frequency scales linearly with the rotation. Indeed we find that the
slope is also proportional to $m$ so that \be {\rm Re}[\omega_{QN}]
\simeq \frac{mBc^2}{A^2}\quad{\rm as}\ \ B \rightarrow
\infty\,,\quad{\rm for}\ \ m>0 \label{mpos} \ee We notice that this
behavior was already present in the WKB investigation in
\cite{waveanalog1}. In Fig. \ref{fig:f2} we show the imaginary part
of the QN frequencies as a function of the rotation parameter, for
$m=1$. Different overtones have different imaginary parts. Note also
that for high $B$ the real part of the modes coalesce whereas the
imaginary part does not. The magnitude of ${\rm Im}[\omega_{QN}]$
increases with $B$, which was observed also in the WKB approach
\cite{waveanalog1}. Thus, as the rotation increases the perturbation
dies off quicker. This also means that the black hole is stable
against $m>0$ perturbations, because the imaginary part is always
negative.

\noindent {\bf (ii) $m<0$:} In Figs. \ref{fig:f3}-\ref{fig:f4} we
show results concerning perturbations having negative $m$, i.e.,
counter-rotating waves.  The behavior of the QN frequencies for
$m<0$ is drastically different from the $m>0$ perturbations.  In
Fig. \ref{fig:f3} we plot the dependence of ${\rm Re}[\omega_{QN}]$
as a function of the rotation of the black hole $B$. As $B$
increases the magnitude of the real part of the QN frequency
decreases.

The oscillation frequencies for the fundamental modes, labeled by
$n=0$, indeed get close to the horizontal axis as $B$ goes to
infinity. However, we haven't been able to track some overtone modes
with negative $m$ for very high rotation since, as can be seen in
Fig. \ref{fig:f3}, the real part of these modes eventually change
sign.  It is extremely difficult, using the method employed here, to
compute modes having ${\rm Re}[\omega_{QN}] \sim 0$. Nevertheless,
supposing that (as the numerical studies for the fundamental modes
indicate) the QN frequencies asymptote to zero for very large B, a
WKB analysis shows that $\omega_{QN} \sim k/B$, where $k$ is some
$m$-dependent constant.  The imaginary part of the QN frequencies
behaves in a similar manner, as seen in Fig. \ref{fig:f4}.

\noindent {\bf (iii) $m=0$:} For circularly symmetric ($m=0$) modes,
our numerical method shows no sign of convergence. For $m=0$, the
wave equation can be written in the simpler form \be \Psi_{,\hat r_*
\hat r_*}+(\omega ^2-V) \Psi=0\,, \label{waveequation3} \ee where
\be V=\left(\f{\hat r^2-1}{\hat r^2}\right) \left[-\f{1}{4\hat
r^2}+\f{5}{4\hat r^4}\right]\,. \ee The potential $V$ is not
positive definite, and this precludes also a simple stability proof.

To have a better physical understanding of this data, consider a
$m=1$ mode, for which the lowest mode (this is the mode that
controls the ringing phase) is approximately $\omega _{QN} \sim
(0.4-0.33i)\frac{c}{r_H}$, with $r_H$ the horizon radius. If one
builds an acoustic black hole by making a $r_H=1\,{\rm mm}$ hole in
a tub with water, then this black hole should have a characteristic
ringing frequency of $\omega \sim 4\times 10^{5}\, {\rm s}^{-1}$,
and a typical damping timescale given by $\tau=\frac{1}{{\rm
Im}[\omega]}\sim 3\times 10^{-6} \,{\rm s}$.
\subsection{Late-time tails}
The existence of late-time tails in black hole spacetimes is well
established, both analytically and numerically, in linearized
perturbations and even in a non-linear evolution, for massless or
massive fields \cite{tails}. This is a problem of more than academic
interest: one knows that a black hole radiates away everything that
it can, by the so called no hair theorem (see \cite{bek} for a nice
review), but how does this hair loss proceed dynamically? A more or
less complete picture is now available. The study of a fairly
general class of initial data evolution shows that the signal can
roughly be divided in three parts: (i) the first part is the prompt
response, at very early times, and the form depends strongly on the
initial conditions. This is the most intuitive phase, being the
obvious counterpart of the light cone propagation. (ii) at
intermediate times the signal is dominated by an exponentially
decaying ringing phase, and its form depends entirely on the black
hole characteristics, through its associated quasinormal modes
\cite{kokkotas,cardososhijunlemos,cardosoAdS}. (iii) a late-time
tail, usually a power law falloff of the field. This power law seems
to be highly independent of the initial data, and seems to persist
even if there is no black hole horizon. In fact it depends only on
the asymptotic far region.

It is not generally appreciated that there is another case in which
wave propagation develops tails: wave propagation in odd dimensional
{\it flat} spacetimes.  In fact, the Green's function in a
$D$-dimensional spacetime (see Cardoso {\it et al} in \cite{tails}
and also \cite{greend,amj}) have a completely different structure
depending on whether $D$ is even or odd. For even $D$ it still has
support only on the light cone, but for odd $D$ the support of the
Green's function extends to the interior of the light cone, and
leads to the appearance of tails.

Analogue black holes also exhibit late-time tails, shedding their
hair in a power-law falloff manner. As an elegant application of the
wave tail formalism developed by Ching {et al} \cite{tails}, it was
found in \cite{waveanalog1} that any perturbation in the vicinity of
the ($2+1$)-dimensional analogue black hole described by
(\ref{metric2}) eventually decays as a power-law falloff of the form
\be H \sim t^{-(2m+1)}\,. \label{latet} \ee
On the other hand, this s precisely the tails that appear in any
($2+1$)-dimensional {\it flat} spacetime (see Cardoso {\it et al} in
\cite{tails}). We thus have a consistent and elegant result.
\subsection{Superradiant amplification of phonons}
Rotating black holes can superradiate, in the sense that in a
scattering experiment the scattered wave has a larger {\it
amplitude} (the frequency is the same, this is {\it not} a Doppler
effect has explained in \cite{zel} and references therein) than the
incident wave. Superradiance is a general phenomenon in physics.
Inertial motion superradiance has long been known \cite{ginzburg},
and refers to the possibility that a (possibly electrically neutral)
object endowed with internal structure, moving uniformly through a
medium, may emit photons even when it starts off in its ground
state. Some examples of inertial motion superradiance include the
Cherenkov effect, the Landau criterion for disappearance of
superfluidity, and Mach shocks for solid objects traveling through a
fluid (cf. \cite{bekenstein} for a discussion). Non-inertial
rotational motion also produces superradiance. This was discovered
by Zel'dovich \cite{zel}, who pointed out that a cylinder made of
absorbing material and rotating around its axis with frequency
$\Omega$ can amplify modes of scalar or electromagnetic radiation of
frequency $\omega$, provided the condition
\be\label{suprad} \omega<m\Omega \ee
(where $m$ is the azimuthal quantum number with respect to the axis
of rotation) is satisfied. Zel'dovich realized that, accounting for
quantum effects, the rotating object should emit spontaneously in
this superradiant regime. He then suggested that a Kerr black hole
whose angular velocity at the horizon is $\Omega$ will show both
amplification and spontaneous emission when the condition
(\ref{suprad}) for superradiance is satisfied. This suggestion was
put on firmer ground by a substantial body of work \cite{superr}. In
particular, it became clear that (even at the purely {\it classical}
level) superradiance is required to satisfy Hawking's area theorem
\cite{beksuperr,teupress}.

Superradiance is essentially related to the presence of an
ergosphere, allowing the extraction of rotational energy from a
black hole through a wave equivalent of the Penrose process
\cite{penrose}. Under certain conditions, superradiance can be used
to induce instabilities in Kerr black holes \cite{schwinger}.
Indeed, all spacetimes admitting an ergosphere and {\it no horizon}
are unstable due to rotational superradiance. This was shown
rigorously in \cite{friedman}, but the growth rate of the
instability is too slow to observe it in an astrophysical context
\cite{cominsschutz,shinichirou}. Kerr black holes are stable, but if
enclosed by a reflecting mirror they can become unstable due to
superradiance \cite{teupress,bhb};

The possibility to observe rotational superradiance in analogue
black holes was considered by Sch\"utzhold and Unruh
\cite{schutzhold}, and more extensively by Basak and Majumdar
\cite{basak1,basak2}, who computed analytically the reflection
coefficients in the low frequency limit $\omega A/c^2\ll 1$. In
particular, the authors of \cite{schutzhold} showed that the
ergoregion instability in gravity wave analogues is related to the
existence of an ``energy function'' [their Eq. (68)] that is not
positive definite inside the ergosphere. In the context of
analogues, inertial superradiance based on superfluid $^3$He has
been studied by Jacobson and Volovik \cite{tedvolovik}. The explicit
numerical calculation of reflection coefficients for the
($2+1$)-rotating acoustic black hole was done in \cite{waveanalog1},
in the superradiant regime.

\begin{figure}
\centerline{\mbox{ \psfig{figure=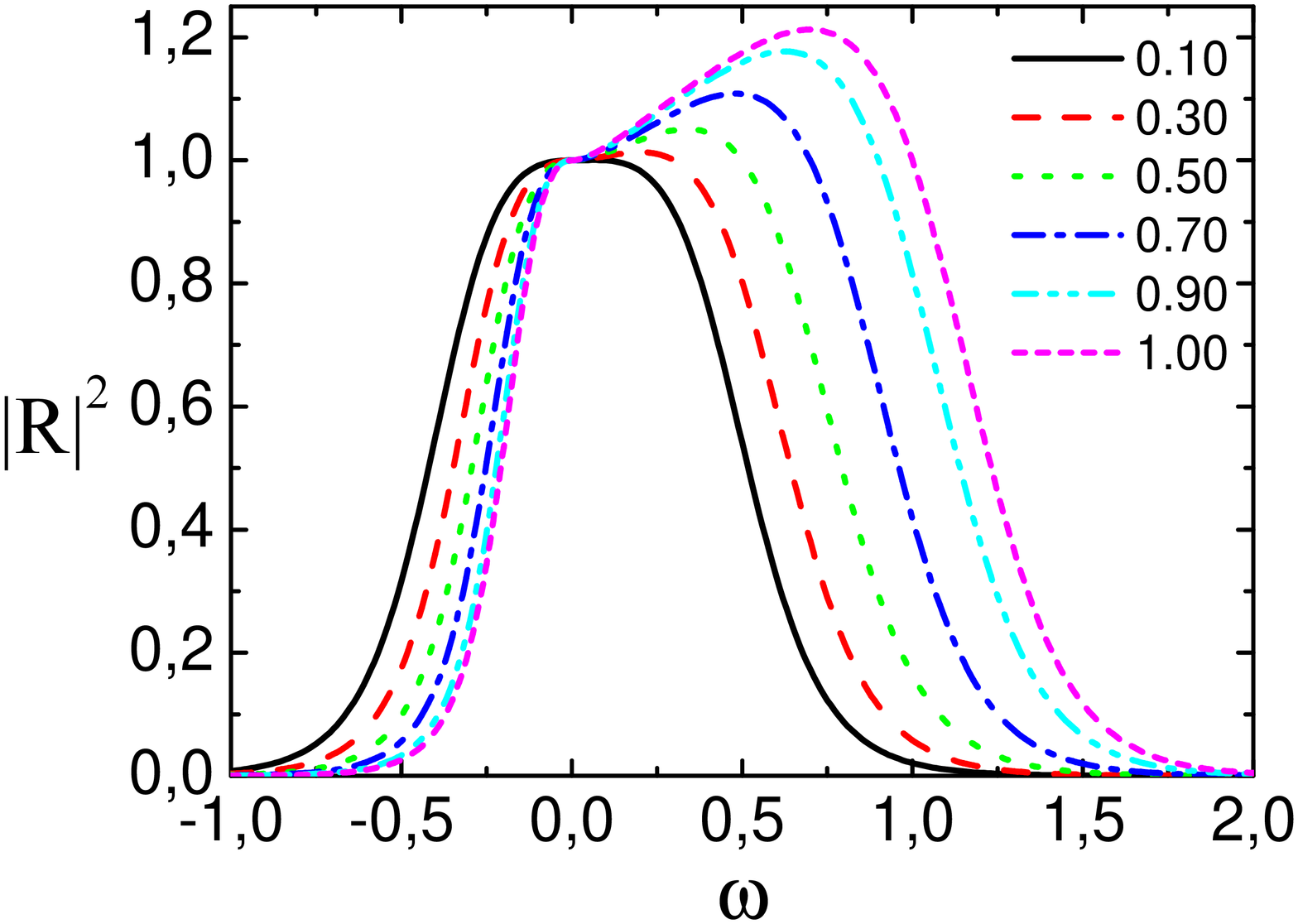,angle=270,width=7cm}
}} \centerline{\mbox{ \psfig{figure=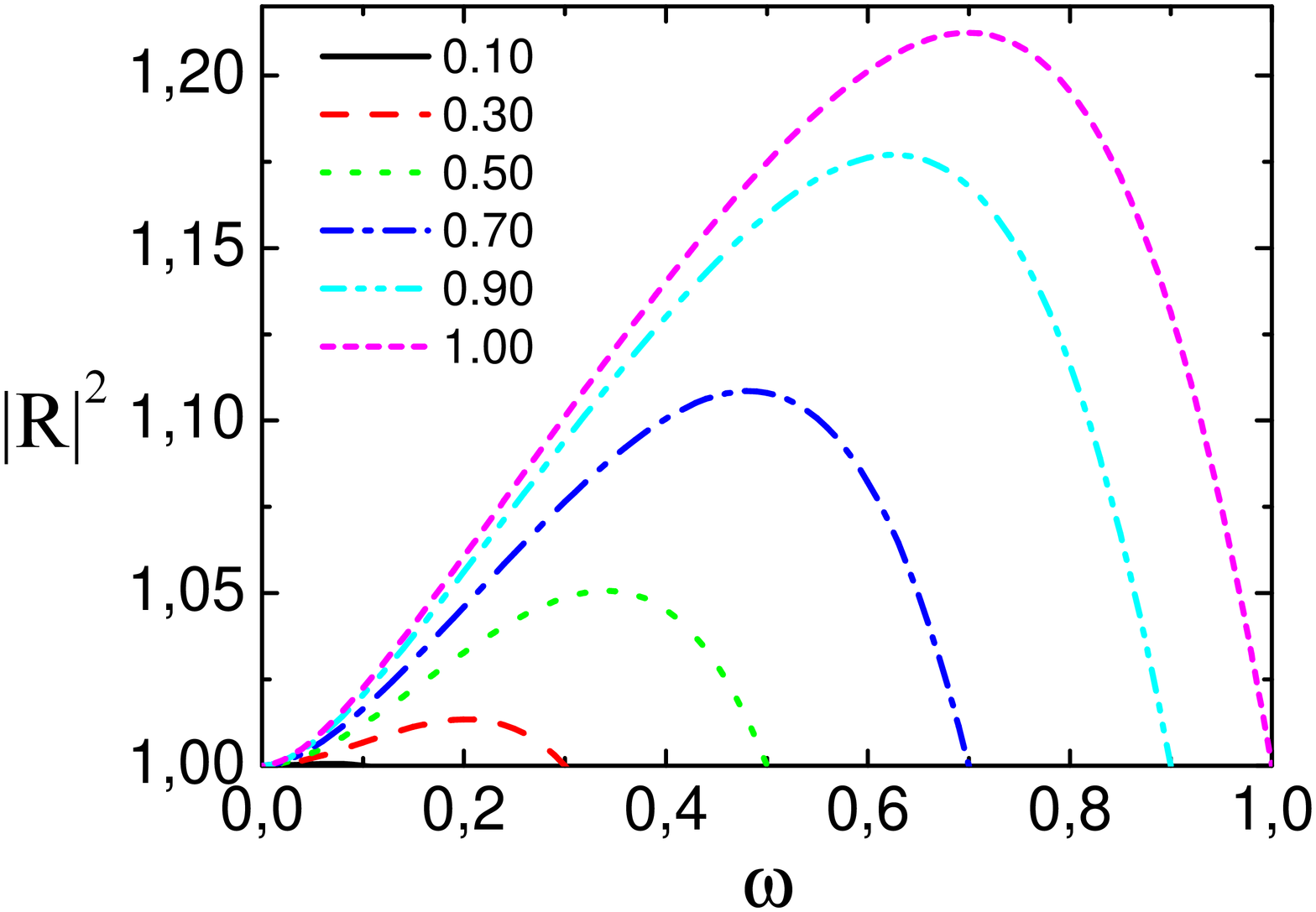,angle=270,width=7cm}
}} \centerline{\mbox{ \psfig{figure=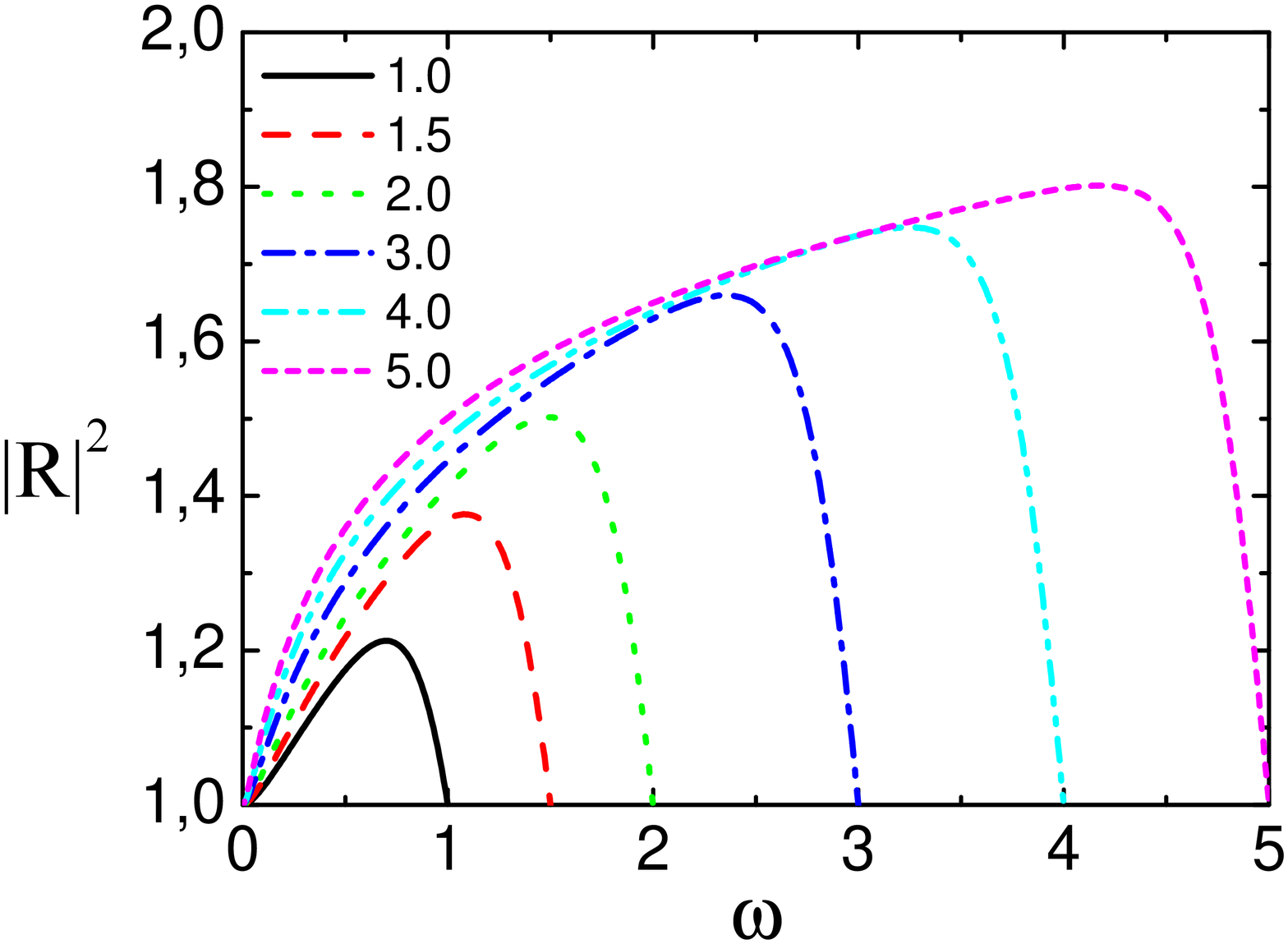,angle=270,width=7cm}
}} \caption{ Reflection coefficient $\left|R_{\omega m}\right|^2$ as
a function of $\omega$ for $m=1$. Each curve corresponds to a
different value of $B$, as indicated. The top panels show that the
reflection coefficient decays exponentially at the critical
frequency for superradiance, $\omega_{SR}=mB$. The middle panels
show a close-up view in the superradiant regime for $B<1$: at $B=1$
the maximum amplification is 21.2 \%. The bottom panels show that
superradiant amplification can become much more efficient for values
of the rotation parameter $B>1$. } \label{fig:superradiantfactor}
\end{figure}

Results of the numerical integrations for the draining bathtub
metric are shown in Fig. \ref{fig:superradiantfactor}. Panels on the
left show the reflection coefficient $|R_{\omega 1}|^2$ for $m=1$,
and panels on the right show $|R_{\omega 2}|^2$ for $m=2$, for
selected values of the black hole rotation $B$. Panels on top show
that, as expected, in the superradiant regime $0<\omega<mB$ the
reflection coefficient $|R_{\omega m}|^2\geq 1$. Furthermore, as one
increases $B$ the reflection coefficient increases, and for fixed
$B$, the reflection coefficient $\left |R_{\omega m}\right|^2$
attains a maximum at $\omega \sim mB$, after which it decays
exponentially as a function of $\omega$ outside the superradiant
interval. This is very similar to what happens when one deals with
massless fields in the vicinities of rotating Kerr black holes
\cite{teupress}. In particular, from the close-up view in the middle
panels we see that, for $B=1$, the maximum amplification is 21.2 \%
($m=1$) and 4.7 \% ($m=2$).

As a final remark, and as we have anticipated, an important
difference between the acoustic black hole metric and the Kerr
metric is that in the present case there is no mathematical upper
limit on the black hole's rotational velocity $B$. In the bottom
panels we show that, considering values of $B>1$, we can indeed have
larger amplification factors for acoustic black holes.

Summarizing: if we are clever enough to build in the lab an acoustic
black hole that spins very rapidly, rotational superradiance can be
particularly efficient in analogues. This is an important result,
considering that the detection of rotational superradiance in the
lab is by no means an easy task, as originally predicted by
Zel'dovich \cite{zel} and confirmed by recent reconsideration of the
problem \cite{bekenstein}. Of course, in any real-world experiment
the maximum rotational parameter will be limited. At the
mathematical level, the equations describing sound propagation
(which are written assuming the hydrodynamic approximation) will
eventually break down. Physically, if the angular component of the
velocity $v^\theta$ becomes very large the dispersion relation for
the fluid will change, invalidating the assumptions under which we
have derived our acoustic metric \cite{schutzhold,waveanalog1}.

The superradiant phenomena we have described are purely {\it
classical} in nature. However, an interesting suggestion to observe
{\it quantum} effects in acoustic superradiance was put forward in
\cite{basak2}. To write down our acoustic metric we required the
flow to be irrotational and non-viscous. As a natural choice, we
could use a fluid which is well known to possess precisely these
properties: superfluid HeII. In this case the presence of vortices
with quantized angular momenta may lead to a quantized energy flux.
The heuristic argument presented in \cite{basak2} goes as follows.
Let us imagine that our black hole is a vortex with a sink at the
center. In the quantum theory of HeII the wavefunction is of the
form $\Psi=\exp\left[\ii \sum_j \phi(\vec r_j) \Phi_{\rm
ground}\right]$, where $\vec r_j$ is the position of the $j$-th
particle of HeII. The velocity at any point is given by the gradient
of the phase at that point, $\vec v=\nabla \phi$, so that (roughly
speaking) the velocity potential can be identified with the phase of
the wavefunction. This phase will be singular at the sink $r=0$.
Continuity of the phase around a circle surrounding the sink
requires that the change of the wavefunction satisfies $\Delta
\phi=2\pi B$. For the wavefunction to be single valued, $B$ (that
is, the black hole's angular velocity at the horizon) must be the
integer multiple of some minimum value $\Delta B$, i.e.,
$B=n\,\Delta B$. Then the angular momentum of the acoustic black
hole would be forced to change in integer multiples of $\Delta B$.
Correspondingly, the spectrum of the reflection coefficients may be
given by equally-spaced peaks with different strengths. This
discrete amplification could enhance chances of observing
superradiance in acoustic black holes, and rule out (or provide
empirical support to) some of the many competing heuristic
approaches to black hole quantization.
\subsection{Acoustic black branes and superradiant instabilities}
Now, we are free to add an extra dimension and interpreting the
result as the superposition of a vortex filament and a line sink
\cite{visser}. We get therefore the following line element \beq
ds^2&=& -\left(1-\frac{A^2+B^2}{c^2r^2} \right)c^2 d\tilde{t}^2+
\left(1-\frac{A^2}{c^2r^2} \right )^{-1}dr^2 -\nonumber \\
& & 2Bd\tilde{\phi}d\tilde{t}+r^2d\tilde{\phi}^2+dz^2\,.
\label{metricbrane} \eeq This describes an analogue black brane, and
compactification of the transverse direction $z$ can be
accomplished, in practice by using a ``tamper'' around the flow.

The propagation of a sound wave in a barotropic inviscid fluid with
irrotational flow is described by the Klein-Gordon equation
$\nabla_{\mu}\nabla^{\mu}\Psi=0$ for a massless field $\Psi$ in a
Lorentzian acoustic geometry, which in our case takes the form
(\ref{metricbrane}). In our acoustic geometry we can separate
variables by the substitution
\be \Psi(\tilde{t},r,\tilde{\phi})=\frac{1}{\sqrt{r}} H(r)e^{\ii
(m\tilde{\phi}+\mu z-\omega \tilde{t})}\,, \ee
and we get the wave equation

\beq H_{,r_* r_*}+ \left\{\left(\omega-\f{Bm}{r^2}\right)^2-
V\right\}
H=0 \\
V=f \left[\f{1}{r^2}\left(m^2-\f{1}{4}\right)+\f{5}{4r^4}+\mu
^2\right] \,. \label{waveequationmass1} \eeq To arrive at
(\ref{waveequationmass1}) we have already performed the following
re-scaling: $\hat r=rA/c$, $\hat{\omega}=\omega A/c^2$,
$\hat{B}=B/A$, $\hat \mu=\mu c/A$. The re-scaling effectively sets
$A=c=1$ in the original wave equation, and picks units such that the
acoustic horizon $\hat r_H=1$. The quantity $f\equiv
(1-\frac{A^2}{c^2r^2})$, and the tortoise coordinate $r_*$ is
defined by the condition
\be \frac{dr_*}{dr}=\frac{1}{f}\,.\ee Explicitly, \be
r_*=r+\f{A}{2c}\log\left|\f{cr-A}{cr+A}\right|\,. \ee

\begin{figure}
\centerline{\mbox{ \psfig{figure=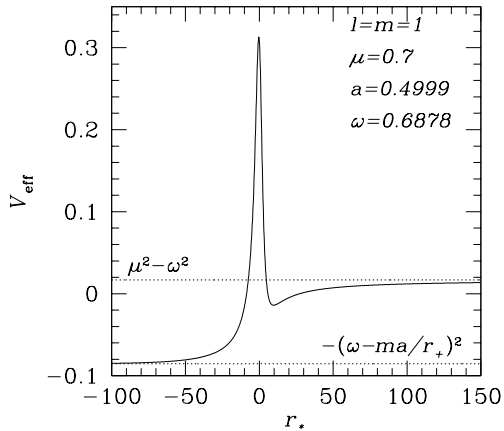,angle=0,width=7cm} }}
\caption{A typical form for the effective potential in the Kerr
geometry, here shown for $l=m=1$ modes. We have set the mass of the
black hole $M=1$, so the rotation parameter $a$ varies between $0$
(Schwarzschild limit) and $1/2$ (extremal limit). Here we plot the
effective potential for the near extreme situation, $a \sim 0.5$ and
for $\mu=0.7$ and $\omega=0.6878$.} \label{fig:fbound}
\end{figure}

It is known that the Kerr geometry, or any rotating (absorbing) body
displays superradiance \cite{zel,misnerunruhbekenstein}. This means
that in a scattering experiment of a wave with frequency
$\omega<m\Omega$ the scattered wave will have a larger amplitude
than the incident wave, the excess energy being withdrawn from the
object's rotational energy. Here $\Omega$ is the horizon's angular
velocity and $m$ is an azimuthal quantum number. Now suppose that
one encloses the rotating black hole in a spherical mirror. Any
initial perturbation will get successively amplified near the black
hole event horizon and reflected back at the mirror, thus creating
an instability. This is the black hole bomb, as devised in
\cite{teupress} and recently improved in \cite{bhb}. This
instability is caused by the mirror, which is an artificial wall,
but one can devise ``natural mirrors'' if one considers massive
fields. Imagine a wavepacket of the massive field in a distant
circular orbit. The gravitational force binds the field and keeps it
from escaping or radiating away to infinity. But at the event
horizon some of the field goes down the black hole, and if the
frequency of the field is in the superradiant region then the field
is amplified. Hence the field is amplified at the event horizon
while being bound away from infinity. Yet another way to understand
this, is to think in terms of wave propagation in an effective
potential. If the effective potential has a well, then waves get
``trapped'' in the well and amplified by superradiance, thus
triggering an instability. In the case of massive fields on a
(four-dimensional) Kerr background, the effective potential indeed
has a well, as we show in Figure \ref{fig:fbound}. Consequently, the
massive field grows exponentially and is unstable (see
\cite{schwinger,detweiler,zouros,furu,braneinstability} for explicit
examples).
With this in mind, we would expect that black strings and branes of
the form (\ref{metricbrane}) for which there is a bound state will
be unstable; here the transverse direction $z$ works as an effective
mass for the sound wave. To get a bound state, one necessary
condition is that the derivative of the potential is positive, at
asymptotic large radial distances (see \cite{braneinstability} for
more details). Now, near infinity, we get \be V_{eff} \sim
\frac{4m^2-1-\mu ^2+8\omega m B}{4r^2}\,,\ee which leads to \be
V'_{eff} \sim \frac{4\mu ^2+1-4m^2-8\omega m B}{2r^3}\,.\ee Now,
this can be positive, thus an instability can be triggered.

For the sake of generality, let us drop the constant $\rho$
requirement, which means not assuming conservation of angular
momentum (this can be achieved by having external torques), then we
can show that the effective metric is
\begin{eqnarray}
ds^2\!\!&=&\!\!
-\left(1-\frac{A^2/\rho ^2+B^2}{c^2r^2} \right)c^2 d\tilde{t}^2  \nonumber \\
& &+ \left(1-\frac{A^2}{\rho ^2c^2r^2} \right )^{-1}dr^2
-2B d\tilde{\phi}d\tilde{t}+r^2d\tilde{\phi}^2+dz^2\,. \nonumber \\
& & \label{metric3}
\end{eqnarray}
Here, $A\,,B$ are again constants but they have different
dimensions. Separating variables by the substitution
\be \Phi(\tilde{t},r,\tilde{\phi})=
 \sqrt{r} \Psi(r)e^{i (\mu z + m\tilde{\phi}-\omega\tilde{t})}\,,
\ee
implies that $\Psi(r)$ obeys the wave equation (just insert the
ansatz in Klein-Gordon's equation)

\be \frac{d^2 \Psi}{dr_* ^2}+\left( (\omega -\frac{Bm}{r^2})^2 -V
\right )\Psi=0\,.\ee

Here \beq V&=& f\left (\mu
^2c+\frac{4m^2c-c}{4r^2}+\frac{c'}{2r}\right )+\nonumber
\\ & & f \left (\frac{A^2}{4cr^4\rho ^2}\left ( 5+2r(\frac{c'}{c}+\frac{2\rho
'}{\rho})\right ) \right )\,,\eeq and the tortoise coordinate $r_*$
is defined as \be \frac{dr}{dr_*}=c(1-\frac{A^2}{c^2\rho ^2
r^2})\equiv f\,.\ee Notice that for constant $\rho\,,\, c$ one
recovers the equations (\ref{waveequationmass1}), as one should.
Now, it is quite easy to present an example flow which the
instability is triggered: take for instance a flow for which $\rho$
is almost constant at infinity (almost means that it asymptotes to a
constant value more rapidly than the sound velocity). Assume also
that, near infinity, $c=c_1+\frac{c_2}{r}$. Then, we get that near
infinity the effective potential behaves as \be
\frac{2c_1c_2k^2}{r}\,.\ee For this to have a positive derivative,
one requires $c_2<0$ ($c_1$ must be positive, as it is the
asymptotic value of the sound velocity). We thus have one example of
flow for which the instability is active. There are many others, of
course, and there are also instances for which the system is stable.
\subsection{Absorption cross-sections}
The computation of absorption cross-sections may be handled
analytically in the low frequency regime, to which I now turn. The
computation of absorption cross-sections of different gravitational
black holes has gained a special interest some years ago, since it
was shown that string theory could reproduce these results (for some
particular geometries). I refer the reader to \cite{oz} for a
introduction to the subject. Considering again our
($2+1$)-dimensional rotating acoustic black hole we shall now
attempt at solving the wave equation in this geometry, in the limit
of small $\omega$. The method we use here follows closely the work
of Starobinsky and Churilov \cite{superr} and Unruh and others
\cite{cross}. The wave equation reads \cite{waveanalog1}: \beq
H_{,r_* r_*}+ \left\{\left(\omega-\f{Bm}{r^2}\right)^2- V\right\}
H=0 \\
V=f \left[\f{1}{r^2}\left(m^2-\f{1}{4}\right)+\f{5}{4r^4}\right] \,.
\label{waveequationmass}\eeq The computation will follow closely
that in \cite{cardosodias}. Changing wavefunction to $R=r^{-1/2}H$
we find that $R$ satisfies \be \Delta
\partial _r \left ( \Delta \partial _r R  \right )+
\left( \omega^2r^2-2Bm\omega+ \frac{B^2m^2}{r^2}-\frac{\Delta
m^2}{r} \right ) R=0\,,\label{waveeqmod} \ee where $\Delta=r-1/r$.
Let us solve this equation in the far-region, $r>>1$. We then have,
near infinity, \be \Delta
\partial _r \left ( \Delta \partial _r R  \right ) \sim r^2 \partial ^2 _r R +r\partial _r R\,.\label{waveapinf} \ee
Assuming $B\omega <<1$, the wave equation (\ref{waveeqmod}) in this
region takes the form \be r^2 \partial ^2 _r R +r\partial _r R +
\left ( \omega ^2 r^2 -m^2\right )R=0 \,.\ee Defining $\rho =\omega
r$ this takes the form \be \rho^2 \partial^2 _{\rho}R +\rho
\partial _{\rho} R +\left ( \rho^2-m^2 \right )R=0\,,\label{farfinal}\ee
which is a Bessel equation (see for example \cite{abramowitz,niki}),
with the general solution \be R=\alpha J_m (\omega r)+\beta Y_m
(\omega r)\,.\ee Notice that $m$ is an integer, and therefore $J_m$
and $J_{-m}$ are not linearly independent.

We will want later on to do a matching between the near region
solution and the far-region solution, so let us investigate the near
region behavior of this solution, or the limit $\omega r\rightarrow
0 $. We find \cite{cardosodias} \be R \sim
\frac{\alpha}{\Gamma[m+1]}(\frac{\omega
r}{2})^{m}-\frac{\beta}{\pi}\Gamma[m](\frac{\omega
r}{2})^{-m}\,,\label{match1}\ee where $\psi$ is the digamma
function.
We now define the near-region as the range for which
$r-r_+<<\frac{1}{\omega}$. Defining \beq z=\frac{r_+}{r^2} \\
R=z^{\alpha} (1-z)^{\beta}T  \,,\eeq then, in this region the
solution representing ingoing waves at the horizon (which is the
boundary condition one must impose) may be written as
\cite{cardosodias} \be T=A_1 F[a,b,a+b-c+1,1-z]\,,\ee where \beq
a=1+\frac{m}{2}-i\varpi\,,
\\ b=\frac{m}{2}-i\varpi\,, \\ c=1+m \,.\eeq Here $\varpi \equiv
\frac{\omega-mB}{2}$, and $F[.,.,.,]$ denotes the standard
hypergeometric functions. Since $c=m+1$ is a integer, one must be
very careful in handling the hypergeometric function
\cite{abramowitz,niki}.

When $z \rightarrow 0$, or $r \rightarrow \infty$, we have \beq
R\sim Ar^{-m} \{ \psi (a) +\psi (b) -\psi (1+m)-\psi (1)\}+ \nonumber\\
Ar^{m}\{ \frac{(-1)^{m-1}\Gamma[m]}{m!(a-m)_m (b-m)_m} \}
\,,\label{match2}\eeq where $A$ is some constant and $(a)_m$ stands
for $(a)_m=a(a+1)(a+2)...(a+m-1)\,,(a_0)=1$. Matching the two
solutions (\ref{match1}) and (\ref{match2}) we get \be
\frac{\alpha}{\beta}=\frac{1}{\pi}\frac{(-1)^{m}\Gamma[m]^2(\omega/2)^{-2m}}{(a-m)_m(b-m)_m
(\psi(a)+\psi(b)-\psi(m+1)-\psi(1))}\,.\ee Now, $(a-m)_m
=\frac{\Gamma[a]}{\Gamma[a-m]}$, we thus have

\be
\frac{\alpha}{\beta}=\frac{1}{\pi}\frac{(-1)^{m}\Gamma[m]^2\Gamma[a-m]\Gamma[b-m](\omega/2)^{-2m}}{\Gamma[a]\Gamma[b]
(\psi(a)+\psi(b)-\psi(m+1)-\psi(1))}\,. \label{csfinal}\ee

This is the final expression. Since $\alpha, \beta$ are related to
the amplitude of in- and out-going waves at infinity, using
(\ref{csfinal}) we can straightforwardly compute reflection
coefficients, absorption cross sections, etc.

\section{Conclusions}
Analogue black holes have proven to be a very valuable tool for the
investigation of problems related to Hawking radiation. It is also
possible that will yield valuable information regarding {\it
classical} phenomena involving black holes. We have shown here some
aspects of classical phenomena involving acoustic black holes, that
may prove useful for future experimental realization of these
systems.

\section*{Acknowledgements}
I would like to take this opportunity to thank Emanuele Berti,
\'Oscar Dias, Jos\'e Lemos, M\'ario Pimenta, Ana Sousa and Shijun
Yoshida for many useful conversations and collaboration. I also
acknowledge financial support from FCT through grant SFRH/BPD/2004.

\end{document}